\documentclass[12pt]{article}
\NewDocumentCommand\citeproctext{}{}

\makeatletter

 \let\@cite@ofmt\@firstofone
 \def\@biblabel#1{}
 \def\@cite#1#2{{#1\if@tempswa , #2\fi}}
\makeatother
\newlength{\cslhangindent}
\newlength{\csllabelwidth}
\newenvironment{CSLReferences}[2] 
 {\begin{list}{}{%
  \setlength{\itemindent}{0pt}
  \setlength{\leftmargin}{0pt}
  \setlength{\parsep}{0pt}
  \ifodd #1
   \setlength{\leftmargin}{\cslhangindent}
   \setlength{\itemindent}{-1\cslhangindent}
  \fi
  \setlength{\itemsep}{#2\baselineskip}}}
 {\end{list}}
\usepackage{calc}

\RequirePackage{titling}
\pretitle{
  \begin{center}
  \bfseries
}
\posttitle{
  \par 
  \vskip 0.5em
  \end{center}
}
\preauthor{ 
  \begin{center}
  \begin{tabular}[t]{c} \normalsize}
\postauthor{\end{tabular}\par
\end{center}
}

\predate{\begin{center}\small}
\postdate{\par\end{center}}

\makeatletter
\renewenvironment{abstract}{
  \small
  \quotation
}
{
  \endquotation
}
\makeatother

\RequirePackage{titlesec}
\titleformat{\section}
  {\bfseries\centering} 
  {\thesection}
  {1em} 
  {}  
\titleformat{\subsection}
  {\bfseries} 
  {\thesubsection}
  {1em} 
  {}  
\titleformat{\subsubsection}
  {} 
  {\thesubsubsection}
  {1em} 
  {}

\RequirePackage[font=small]{caption}

\RequirePackage[]{geometry}

\RequirePackage[T1]{fontenc}

\RequirePackage{amsmath,amsthm}
\RequirePackage{amssymb}

\RequirePackage[colorlinks=true,
                linkcolor=blue,
                urlcolor=black,
                citecolor=black]{hyperref}

\RequirePackage{cleveref}

\RequirePackage{longtable,booktabs,array}
\usepackage{calc} 
\usepackage{etoolbox}
\makeatletter
\patchcmd\longtable{\par}{\if@noskipsec\mbox{}\fi\par}{}{}
\makeatother
\IfFileExists{footnotehyper.sty}{\usepackage{footnotehyper}}{\usepackage{footnote}}
\makesavenoteenv{longtable}

\RequirePackage{color}
\RequirePackage{braket}
\RequirePackage{graphicx}
\RequirePackage{comment}
\RequirePackage[shortlabels,inline]{enumitem}
\RequirePackage{float}

\RequirePackage{tikz}
\RequirePackage{tikz-3dplot}

\usetikzlibrary{patterns, fit, decorations.pathreplacing, shadings}
\usetikzlibrary{trees, arrows}

\input{preamble.tex}

\newunicodechar{ℍ}{\ensuremath{\mathbb{H}}}

\newunicodechar{ℕ}{\ensuremath{\mathbb{N}}}

\newunicodechar{ℙ}{\ensuremath{\mathbb{P}}}
\newunicodechar{ℚ}{\ensuremath{\mathbb{Q}}}
\newunicodechar{ℝ}{\ensuremath{\mathbb{R}}}

\newunicodechar{ℤ}{\ensuremath{\mathbb{Z}}}

\newunicodechar{ℓ}{\ensuremath{\ell}}
\newunicodechar{α}{\ensuremath{\alpha}}
\newunicodechar{β}{\ensuremath{\beta}}
\newunicodechar{γ}{\ensuremath{\gamma}}
\newunicodechar{Γ}{\ensuremath{\Gamma}}
\newunicodechar{δ}{\ensuremath{\delta}}
\newunicodechar{Δ}{\ensuremath{\Delta}}
\newunicodechar{ϵ}{\ensuremath{\epsilon}}
\newunicodechar{ζ}{\ensuremath{\zeta}}
\newunicodechar{η}{\ensuremath{\eta}}
\newunicodechar{∑}{\ensuremath{\sum}}
\newunicodechar{θ}{\ensuremath{\theta}}
\newunicodechar{Θ}{\ensuremath{\Theta}}
\newunicodechar{ι}{\ensuremath{\iota}}
\newunicodechar{κ}{\ensuremath{\kappa}}
\newunicodechar{λ}{\ensuremath{\lambda}}
\newunicodechar{Λ}{\ensuremath{\Lambda}}
\newunicodechar{μ}{\ensuremath{\mu}}
\newunicodechar{ν}{\ensuremath{\nu}}
\newunicodechar{ξ}{\ensuremath{\xi}}
\newunicodechar{Ξ}{\ensuremath{\Xi}}

\newunicodechar{π}{\ensuremath{\pi}}
\newunicodechar{Π}{\ensuremath{\Pi}}
\newunicodechar{ρ}{\ensuremath{\rho}}
\newunicodechar{𝔼}{\ensuremath{\mathbb{E}}}

\newunicodechar{σ}{\ensuremath{\sigma}}
\newunicodechar{Σ}{\ensuremath{\Sigma}}
\newunicodechar{τ}{\ensuremath{\tau}}

\newunicodechar{υ}{\ensuremath{\upsilon}}
\newunicodechar{Υ}{\ensuremath{\Upsilon}}
\newunicodechar{ϕ}{\ensuremath{\phi}}
\newunicodechar{Φ}{\ensuremath{\Phi}}
\newunicodechar{χ}{\ensuremath{\chi}}

\newunicodechar{ψ}{\ensuremath{\psi}}
\newunicodechar{Ψ}{\ensuremath{\Psi}}
\newunicodechar{ω}{\ensuremath{\omega}}
\newunicodechar{Ω}{\ensuremath{\Omega}}

\newunicodechar{ε}{\ensuremath{\varepsilon}}
\newunicodechar{ϑ}{\ensuremath{\vartheta}}
\newunicodechar{ϰ}{\ensuremath{\varkappa}}
\newunicodechar{ϖ}{\ensuremath{\varpi}}
\newunicodechar{ς}{\ensuremath{\varsigma}}
\newunicodechar{φ}{\ensuremath{\varphi}}

\newunicodechar{ₐ}{\ensuremath{{}_{a}}}
\newunicodechar{ₑ}{\ensuremath{{}_{e}}}
\newunicodechar{ₕ}{\ensuremath{{}_{h}}}
\newunicodechar{ᵢ}{\ensuremath{{}_{i}}}
\newunicodechar{ⱼ}{\ensuremath{{}_{j}}}
\newunicodechar{ₖ}{\ensuremath{{}_{k}}}
\newunicodechar{ₗ}{\ensuremath{{}_{l}}}
\newunicodechar{ₘ}{\ensuremath{{}_{m}}}
\newunicodechar{ₙ}{\ensuremath{{}_{n}}}
\newunicodechar{ₒ}{\ensuremath{{}_{o}}}
\newunicodechar{ₚ}{\ensuremath{{}_{p}}}
\newunicodechar{ᵣ}{\ensuremath{{}_{r}}}
\newunicodechar{ₛ}{\ensuremath{{}_{s}}}
\newunicodechar{ₜ}{\ensuremath{{}_{t}}}
\newunicodechar{ᵤ}{\ensuremath{{}_{u}}}
\newunicodechar{ᵥ}{\ensuremath{{}_{v}}}
\newunicodechar{ₓ}{\ensuremath{{}_{x}}}

\newunicodechar{₀}{\ensuremath{{}_{0}}}
\newunicodechar{₁}{\ensuremath{{}_{1}}}
\newunicodechar{₂}{\ensuremath{{}_{2}}}
\newunicodechar{₃}{\ensuremath{{}_{3}}}
\newunicodechar{₄}{\ensuremath{{}_{4}}}
\newunicodechar{₅}{\ensuremath{{}_{5}}}
\newunicodechar{₆}{\ensuremath{{}_{6}}}
\newunicodechar{₇}{\ensuremath{{}_{7}}}
\newunicodechar{₈}{\ensuremath{{}_{8}}}
\newunicodechar{₉}{\ensuremath{{}_{9}}}

\newunicodechar{₊}{\ensuremath{{}_{+}}}
\newunicodechar{₋}{\ensuremath{{}_{-}}}
\newunicodechar{₌}{\ensuremath{{}_{=}}}

\newunicodechar{ᴬ}{\ensuremath{{}^{A}}}
\newunicodechar{ᴮ}{\ensuremath{{}^{B}}}
\newunicodechar{ᴰ}{\ensuremath{{}^{D}}}
\newunicodechar{ᴱ}{\ensuremath{{}^{E}}}
\newunicodechar{ᴳ}{\ensuremath{{}^{G}}}
\newunicodechar{ᴴ}{\ensuremath{{}^{H}}}
\newunicodechar{ᴵ}{\ensuremath{{}^{I}}}
\newunicodechar{ᴶ}{\ensuremath{{}^{J}}}
\newunicodechar{ᴷ}{\ensuremath{{}^{K}}}
\newunicodechar{ᴸ}{\ensuremath{{}^{L}}}
\newunicodechar{ᴹ}{\ensuremath{{}^{M}}}
\newunicodechar{ᴺ}{\ensuremath{{}^{N}}}
\newunicodechar{ᴼ}{\ensuremath{{}^{O}}}
\newunicodechar{ᴾ}{\ensuremath{{}^{P}}}
\newunicodechar{ᴿ}{\ensuremath{{}^{R}}}
\newunicodechar{ᵀ}{\ensuremath{{}^{T}}}
\newunicodechar{ᵁ}{\ensuremath{{}^{U}}}
\newunicodechar{ⱽ}{\ensuremath{{}^{V}}}
\newunicodechar{ᵂ}{\ensuremath{{}^{W}}}

\newunicodechar{ᵃ}{\ensuremath{{}^{a}}}
\newunicodechar{ᵇ}{\ensuremath{{}^{b}}}
\newunicodechar{ᶜ}{\ensuremath{{}^{c}}}
\newunicodechar{ᵈ}{\ensuremath{{}^{d}}}
\newunicodechar{ᵉ}{\ensuremath{{}^{e}}}
\newunicodechar{ᶠ}{\ensuremath{{}^{f}}}
\newunicodechar{ᵍ}{\ensuremath{{}^{g}}}
\newunicodechar{ʰ}{\ensuremath{{}^{h}}}
\newunicodechar{ⁱ}{\ensuremath{{}^{i}}}
\newunicodechar{ʲ}{\ensuremath{{}^{j}}}
\newunicodechar{ᵏ}{\ensuremath{{}^{k}}}
\newunicodechar{ˡ}{\ensuremath{{}^{l}}}
\newunicodechar{ᵐ}{\ensuremath{{}^{m}}}
\newunicodechar{ⁿ}{\ensuremath{{}^{n}}}
\newunicodechar{ᵒ}{\ensuremath{{}^{o}}}
\newunicodechar{ᵖ}{\ensuremath{{}^{p}}}
\newunicodechar{ʳ}{\ensuremath{{}^{r}}}
\newunicodechar{ˢ}{\ensuremath{{}^{s}}}
\newunicodechar{ᵗ}{\ensuremath{{}^{t}}}
\newunicodechar{ᵘ}{\ensuremath{{}^{u}}}
\newunicodechar{ᵛ}{\ensuremath{{}^{v}}}
\newunicodechar{ʷ}{\ensuremath{{}^{w}}}
\newunicodechar{ˣ}{\ensuremath{{}^{x}}}
\newunicodechar{ʸ}{\ensuremath{{}^{y}}}
\newunicodechar{ᶻ}{\ensuremath{{}^{z}}}

\newunicodechar{⁰}{\ensuremath{{}^{0}}}
\newunicodechar{⁴}{\ensuremath{{}^{4}}}
\newunicodechar{⁵}{\ensuremath{{}^{5}}}
\newunicodechar{⁶}{\ensuremath{{}^{6}}}
\newunicodechar{⁷}{\ensuremath{{}^{7}}}
\newunicodechar{⁸}{\ensuremath{{}^{8}}}
\newunicodechar{⁹}{\ensuremath{{}^{9}}}

\newunicodechar{⁺}{\ensuremath{{}^{+}}}
\newunicodechar{⁻}{\ensuremath{{}^{-}}}
\newunicodechar{⁼}{\ensuremath{{}^{=}}}

\newunicodechar{…}{\ensuremath{\ldots}}
\newunicodechar{⋯}{\ensuremath{\cdots}}
\newunicodechar{⋮}{\ensuremath{\vdots}}

\newunicodechar{–}{\ensuremath{\text{--}}}
\newunicodechar{—}{\ensuremath{\text{---}}}

\newunicodechar{◻}{\ensuremath{\square}}

\newunicodechar{⦅}{\ensuremath{(\!|}}
\newunicodechar{⦆}{\ensuremath{|\!)}}
\newunicodechar{⟨}{\ensuremath{\langle}}
\newunicodechar{⟩}{\ensuremath{\rangle}}
\newunicodechar{⦃}{\ensuremath{\{\!\mid}}
\newunicodechar{⦄}{\ensuremath{\mid\!\}}}
\newunicodechar{⟪}{\ensuremath{\langle\!\langle}}
\newunicodechar{⟫}{\ensuremath{\rangle\!\rangle}}
\newunicodechar{⁅}{\ensuremath{\{}}
\newunicodechar{⁆}{\ensuremath{\}}}
\newunicodechar{｛}{\ensuremath{\{}}
\newunicodechar{｝}{\ensuremath{\}}}

\newunicodechar{⟦}{\ensuremath{[\![}}
\newunicodechar{⟧}{\ensuremath{]\!]}}

\newunicodechar{⌜}{\ensuremath{\ulcorner}}
\newunicodechar{⌝}{\ensuremath{\urcorner}}
\newunicodechar{⌞}{\ensuremath{\llcorner}}
\newunicodechar{⌟}{\ensuremath{\lrcorner}}
\newunicodechar{⌈}{\ensuremath{\lceil}}
\newunicodechar{⌉}{\ensuremath{\rceil}}
\newunicodechar{⌊}{\ensuremath{\lfloor}}
\newunicodechar{⌋}{\ensuremath{\rfloor}}

\newunicodechar{≡}{\ensuremath{\equiv}}
\newunicodechar{¬}{\ensuremath{\lnot}}
\newunicodechar{≢}{\ensuremath{\not\equiv}}
\newunicodechar{∨}{\ensuremath{\lor}}
\newunicodechar{∧}{\ensuremath{\land}}
\newunicodechar{⇒}{\ensuremath{\;\Rightarrow\;}}
\newunicodechar{⇐}{\ensuremath{\;\Leftarrow\;}}
\newunicodechar{⟺}{\ensuremath{\iff}}
\newunicodechar{⇔}{\ensuremath{\iff}}

\newunicodechar{∀}{\ensuremath{\forall}}
\newunicodechar{∃}{\ensuremath{\exists}}

\newunicodechar{❙}{\ensuremath{\,\mid\,}}
\newunicodechar{•}{\ensuremath{\,\bullet\,}}

\newunicodechar{⊢}{\ensuremath{\vdash}}
\newunicodechar{⊣}{\ensuremath{\dashv}}
\newunicodechar{⊨}{\ensuremath{\vDash}}

\newunicodechar{ø}{\ensuremath{\emptyset}}
\newunicodechar{∅}{\ensuremath{\emptyset}}
\newunicodechar{∈}{\ensuremath{\in}}
\newunicodechar{∉}{\ensuremath{\not\in}}
\newunicodechar{∋}{\ensuremath{\ni}}
\newunicodechar{⊆}{\ensuremath{\subseteq}}
\newunicodechar{∩}{\ensuremath{\cap}}
\newunicodechar{∪}{\ensuremath{\cup}}
\newunicodechar{⊍}{\ensuremath{\uplus}}
\newunicodechar{⊎}{\ensuremath{\uplus}}

\newunicodechar{⊥}{\ensuremath{\bot}}
\newunicodechar{⊔}{\ensuremath{\sqcup}}
\newunicodechar{⊓}{\ensuremath{\sqcap}}
\newunicodechar{╲}{\ensuremath{\backslash}}

\newunicodechar{∘}{\ensuremath{\circ}}
\newunicodechar{↦}{\ensuremath{\mapsto}}
\newunicodechar{↪}{\ensuremath{\hookrightarrow}}

\newunicodechar{≠}{\ensuremath{\neq}}
\newunicodechar{≐}{\ensuremath{\doteq}}
\newunicodechar{≟}{\ensuremath{\stackrel{?}{=}}}
\newunicodechar{≅}{\ensuremath{\cong}}
\newunicodechar{≇}{\ensuremath{\ncong}}
\newunicodechar{≃}{\ensuremath{\simeq}}
\newunicodechar{≄}{\ensuremath{\not\simeq}}
\newunicodechar{≈}{\ensuremath{\approx}}
\newunicodechar{≉}{\ensuremath{\not\approx}}
\newunicodechar{∼}{\ensuremath{\sim}}
\newunicodechar{≁}{\ensuremath{\not\sim}}
\newunicodechar{≔}{\ensuremath{:\!=}}

\newunicodechar{≤}{\ensuremath{\leq}}
\newunicodechar{≰}{\ensuremath{\nleq}}
\newunicodechar{≥}{\ensuremath{\geq}}
\newunicodechar{≱}{\ensuremath{\ngeq}}
\newunicodechar{≮}{\ensuremath{\nless}}
\newunicodechar{≯}{\ensuremath{\ngtr}}
\newunicodechar{≦}{\ensuremath{\leqq}}
\newunicodechar{≨}{\ensuremath{\lneqq}}
\newunicodechar{≧}{\ensuremath{\geqq}}
\newunicodechar{≩}{\ensuremath{\gneqq}}
\newunicodechar{≲}{\ensuremath{\lesssim}}
\newunicodechar{≳}{\ensuremath{\gtrsim}}
\newunicodechar{⊏}{\ensuremath{\sqsubset}}
\newunicodechar{⊑}{\ensuremath{\sqsubseteq}}
\newunicodechar{⊐}{\ensuremath{\sqsupset}}
\newunicodechar{⊒}{\ensuremath{\sqsupseteq}}

\newunicodechar{∣}{\ensuremath{\mid}}

\newunicodechar{→}{\ensuremath{\rightarrow}}
\newunicodechar{⇨}{\ensuremath{\rightarrow}}
\newunicodechar{➩}{\ensuremath{\rightarrow}}
\newunicodechar{⟿}{\ensuremath{\rightsquigarrow}}
\newunicodechar{←}{\ensuremath{\leftarrow}}
\newunicodechar{↔}{\ensuremath{\leftrightarrow}}
\newunicodechar{↑}{\ensuremath{\uparrow}}
\newunicodechar{↓}{\ensuremath{\downarrow}}
\newunicodechar{⇊}{\ensuremath{\downarrow\downarrow}}
\newunicodechar{⟶}{\ensuremath{\longrightarrow}}
\newunicodechar{⟵}{\ensuremath{\longleftarrow}}

\newunicodechar{⊕}{\ensuremath{\oplus}}
\newunicodechar{⨁}{\ensuremath{\bigoplus}}
\newunicodechar{⊖}{\ensuremath{\ominus}}
\newunicodechar{⊗}{\ensuremath{\otimes}}
\newunicodechar{⨂}{\ensuremath{\bigotimes}}
\newunicodechar{⊘}{\ensuremath{\oslash}}
\newunicodechar{⊙}{\ensuremath{\odot}}
\newunicodechar{⨀}{\ensuremath{\bigodot}}
\newunicodechar{⊚}{\ensuremath{\circledcirc}}
\newunicodechar{⊛}{\ensuremath{\circledast}}
\newunicodechar{⊝}{\ensuremath{\circleddash}}

\newunicodechar{∙}{\ensuremath{\cdot}}
\newunicodechar{∞}{\ensuremath{\infty}}

\newunicodechar{✔}{\ensuremath{\checkmark}}
\newunicodechar{✓}{\ensuremath{\checkmark}}

\newunicodechar{□}{\ensuremath{QED}}
\newunicodechar{∎}{\ensuremath{QED}}
\newunicodechar{○}{\ensuremath{\circ}}

\newunicodechar{♯}{\ensuremath{SHARP}}

\usepackage{iftex}
\ifPDFTeX
  \usepackage[T1]{fontenc}
  \usepackage[utf8]{inputenc}
  \usepackage{textcomp} 
\else 
  \usepackage{unicode-math} 
  \defaultfontfeatures{Scale=MatchLowercase}
  \defaultfontfeatures[\rmfamily]{Ligatures=TeX,Scale=1}
\fi
\usepackage{lmodern}

\makeatletter
\newsavebox\pandoc@box
\newcommand*\pandocbounded[1]{
  \sbox\pandoc@box{#1}%
  \Gscale@div\@tempa{\textheight}{\dimexpr\ht\pandoc@box+\dp\pandoc@box\relax}%
  \Gscale@div\@tempb{\linewidth}{\wd\pandoc@box}%
  \ifdim\@tempb\p@<\@tempa\p@\let\@tempa\@tempb\fi
  \ifdim\@tempa\p@<\p@\scalebox{\@tempa}{\usebox\pandoc@box}%
  \else\usebox{\pandoc@box}%
  \fi%
}
\renewcommand{\pandocbounded}[1]{#1}

\def\fps@figure{htbp}
\makeatother

\providecommand{\tightlist}{%
  \setlength{\itemsep}{0pt}\setlength{\parskip}{0pt}}

\renewcommand{\tightlist}{}

\usepackage{bookmark}
\IfFileExists{xurl.sty}{\usepackage{xurl}}{} 
\hypersetup{
  pdftitle={Irrelevance of personalized pricing under strategic market segmentation},
  pdfauthor={Xiaoxiao Hu; Haoran Lei},
  hidelinks,
  pdfcreator={LaTeX via pandoc}}

\title{Irrelevance of personalized pricing under strategic market
segmentation}
\author{Xiaoxiao Hu \and Haoran Lei}
\date{\today}

\begin{document}
\newgeometry{margin=1in} 

\maketitle

\newcommand{\cS}{\mathcal{S}}
\newcommand{\ps}{p_2}
\newcommand{\qcav}{\mathrm{qcav}}
\newcommand{\cav}{\mathrm{cav}\,}
\newcommand{\pmap}{\tilde{p}}
\newcommand{\ls}{l_2}
\newcommand{\lm}{l_1}
\newcommand{\cs}{c_2}
\newcommand{\cm}{c_1}
\newcommand{\pro}{\mu}
\newcommand{\an}{a_0}
\newcommand{\cn}{c_0}
\newcommand{\am}{a_1}
\newcommand{\as}{a_2}
\newcommand{\pim}{\pi_1}
\newcommand{\pis}{\pi_2}
\newcommand{\h}{\theta}
\newcommand{\qcpi}{\tilde{\pi}}
\newcommand{\qcv}{\tilde{v}}
\newcommand{\bv}{\bar{v}}
\newcommand{\tv}{\tilde{v}}
\newcommand{\bpi}{\bar{\pi}}
\newcommand{\bp}{\bar{p}}
\newcommand{\hpi}{\hat{\pi}}
\newcommand{\cq}{\hat{q}}
\newcommand{\us}{\underline{s}}
\newcommand{\m}{\mathrm{m}}
\newcommand{\E}{\mathbb{E}}
\newcommand{\Prior}{\mu_0}
\newcommand{\prior}{\mu_0}
\newcommand{\post}{\mu}
\newcommand{\Post}{\mu}
\newcommand{\belief}{\mu}
\newcommand{\EE}{\mathbb{E}}
\newcommand{\cP}{\mathcal{P}}
\newcommand{\katexComment}[1]{\,}
\newcommand{\quasi}{quasiconcavification}
\newcommand{\Quasi}{Quasiconcavification}
\newcommand{\R}{\mathbb{R}}
\newcommand{\co}{\text{co}\,}
\newcommand{\I}{\mathcal{I}}
\newcommand{\ip}{\iota}
\newcommand{\lr}{LR}
\newcommand{\hv}{\hat{v}}
\newcommand{\hq}{\hat{q}}
\newcommand{\seq}[2]{\{ {#1} , \dots ,  {#2} \}}
\newcommand{\GF}{\mathcal{G}^\text{FLW}}
\newcommand{\GA}{\mathcal{G}^\text{A}}
\newcommand{\GH}{\mathcal{G}^\text{HL}}
\newcommand{\bi}{\bar{\imath}}
\newcommand{\ui}{\underline{i}}
\newcommand{\tp}{\tilde{p}}
\newcommand{\FLW}{\text{FLW}}
\newcommand{\qc}{\mathrm{qcav}\,}
\newcommand{\supp}{\mathrm{supp}\,}

\begin{abstract}

A multiproduct seller is more informed than consumers about the value of
her products to consumers. The seller posts a price list and segments
the market through cheap-talk communication. We find that when both
seller's and consumers' incentive-compatibility constraints are
satisfied, the seller cannot benefit from personalized pricing (i.e.,
third-degree price discrimination). Based on that observation, we
provide a tractable characterization of seller's maximum equilibrium
profits. We apply our analysis to a credence-good setup and discuss when
the credence goods seller benefits from communication. The irrelevance
result breaks down when we relax seller's incentive-compatibility
constraints.



\end{abstract}

\thx{

This paper subsumes and generalizes part of the results of a previously
circulated working paper under the title ``Information transmission in
monopolistic credence goods markets.''

Xiaoxiao Hu, Ningbo University. Email: xhuah@connect.ust.hk

Haoran Lei, Hunan University. Email: hleiaa@connect.ust.hk

}

\restoregeometry

\section{Introduction}\label{sec:intro}

In many markets, sellers often possess superior knowledge about the
value of their products or services than consumers do. This is
particularly true for experience goods (such as hotels and second-hand
cars) and credence goods (such as financial advice and taxi rides). With
the growth of digital economy, online platforms like Amazon and Taobao
have amassed vast amounts of consumer data, which could further amplify
sellers' information advantages. These informed sellers often act as
(possibly informal) experts, assisting consumers with their purchasing
decisions. For example, online platforms provide personalized product
recommendations based on consumer data and their own product
information. Additionally, professionals such as lawyers and financial
planners frequently provide free consultations, informing potential
clients about the value of their services.

Motivated by these phenomena, we study a model in which an informed,
multiproduct monopolist strategically reveals product information to
potential consumers through cheap talk (Crawford and Sobel 1982; Green
and Stokey 2007). A typical situation we have in mind is a salesperson
(or online seller) selectively making targeted product recommendations
for different consumers. Intuitively, the monopolist is motivated to
share product information because, by helping consumers make better
purchasing decisions, the monopolist creates extra surplus that can be
(at least partially) captured through higher prices.

Cheap talk means the seller cannot commit to any communication strategy
or provide any hard evidence. Two kinds of equilibria exist where cheap
talk can be influential. In the first one, the seller sets prices to
equalize the marginal profits across a set of products. This allows the
seller to truthfully recommend the best-matching products for consumers,
and consumers willingly follow these recommendations. In the second one,
products have different marginal profits, making the truthful equilibria
unsustainable because the seller would otherwise have an incentive to
recommend the highest-margin product regardless of its fit. To see why
cheap talk can still be influential in this case, consider a concrete
example where a salesperson offers two products.\footnote{The intuitions
  we provide here draw from both cheap talk and credence goods
  literature. Recent cheap talk papers have studied the influential
  equilibria where the sender's (i.e., seller's) payoffs depend solely
  on the receiver's (i.e., buyer's) actions (Chakraborty and Harbaugh
  2010; Lipnowski and Ravid 2020). Lipnowski and Ravid (2020) describe
  the seller's credibility-building strategy as ``degrading self-serving
  information.'' On the other hand, many credence goods papers (e.g.,
  Fong, Liu, and Wright (2014), Fong et al. (2020)) have established
  that buyers must sometimes reject high-margin recommendations to
  ensure credible advice. While credence-good models usually do not
  explicitly use the cheap-talk setup, they work similarly with the
  cheap-talk setup as long as the product quality is verifiable by
  consumers. In \Cref{sec:applications}, we demonstrate an equivalence
  between a seller's optimal equilibrium in the cheap-talk setup and
  that in a credence-good setup.} The salesperson privately knows each
product's value to the buyer, and it is common knowledge that the
products' marginal profits differ. To ensure credible recommendations,
the buyer must occasionally reject the higher-margin product
recommendation, making the salesperson indifferent between recommending
either product. Additionally, to keep the buyer indifferent between
accepting or rejecting the higher-margin product, the salesperson must
still occasionally recommend it when the lower-margin one is a better
match.

When the seller uses strategic communication, consumers are divided into
different segments based on their posterior beliefs about product
values. Intuitively, personalized pricing (i.e., third-degree price
discrimination) seems appealing for the seller, as it allows her to set
different prices for different segments. Without personalized pricing,
the seller must choose a uniform price list across all segments and
ensure consumers choose the target product of their segment; that is,
the seller's optimization problem entails \emph{consumers'
incentive-compatibity constraints.} Personalized pricing avoids these
constraints as the seller can set prohibitively high prices for the
non-targeted products in each segment.

However, we find that personalized pricing generally cannot increase
seller profits compared to uniform pricing (\Cref{prp:main}). Our model
imposes no specific assumptions about the consumer's value function for
products, the seller's cost function, or the players' prior belief.
Therefore, this finding applies broadly to scenarios where the seller
cannot commit to a communication strategy but can steer consumers toward
their target products through cheap talk. The key to our proof of
\Cref{prp:main} is to exploit the seller's incentive-compatibity
constraints that she must be indifferent between assigning a consumer to
any market segment. In other words, we demonstrate that the seller's
incentive-compatibility constraints imposed by cheap talk lead to the
irrelevance of personalized pricing to seller profits.

The irrelevance result also allows us to provide a tractable
characterization of the seller's highest equilibrium profits using the
geometric method by Lipnowski and Ravid (2020). The calculation is
straightforward because we can ignore consumers' incentive constraints.
However, this simplification depends on seller's incentive-compatibility
constraints imposed by cheap talk. If the seller can commit to her
communication strategy, calculating seller's highest profits without
personalized pricing becomes significantly more complicated because we
must account for consumers' incentive constraints, as shown in
\Cref{sec:discuss-persuasion}.

We apply our analysis to a credence-good setting, where the expert
seller makes a product recommendation and the consumer chooses to accept
or reject. We show that the seller's maximum profits with and without
personalized pricing of our main model serve as upper and lower bounds
for credence goods seller profits, respectively. Therefore, our
irrelevance result implies that our profit characterization applies
directly to the credence-good setting. We also provide a necessary and
sufficient condition under which communication benefits the credence
goods seller.

With the rise of digital economy, personalized pricing has become
buzzword and been used in everyday practice. Our finding that informed
sellers do not benefit from personalized pricing might seem unrealistic.
In \Cref{sec:discuss}, we emphasize that the key to understanding this
finding lies in the seller's incentive-compatibility constraints, and
discuss why relaxing these constraints can make personalized pricing
profitable for the seller.

\section{Literature}\label{sec:lit}

This paper is most related to the rapidly growing literature of
``monopoly information design'' (e.g., Lewis and Sappington (1994),
Bergemann, Brooks, and Morris (2015), Roesler and Szentes (2017)), which
explores the role of information disclosure and/or market segmentation
in determining the market outcomes. To the best of our knowledge, Lewis
and Sappington (1994) is the first paper that studies an informed
seller's joint problem of information disclosure and optimal pricing.
Below, we discuss several papers most relevant to our work.

Ichihashi (2020) analyzes a setting where consumers are privately
informed of their valuations and discloses information to the seller.
Ichihashi (2020) finds that it is generally seller-optimal to commit to
non-personalized pricing to encourage more consumer information
disclosure. In our model, the information asymmetry is reversed: the
seller is more informed of the value of her products and discloses
information to consumers through cheap talk. We show that personalized
pricing still does not benefit the seller, since uniform pricing yields
the same equilibrium profits as personalized pricing.

Bergemann, Heumann, and Morris (2022) study a multiproduct seller who
controls both information disclosure and the selling mechanism. In their
model, the seller commits to some communication strategy and then solves
a static screening problem (Mussa and Rosen 1978; Maskin and Riley
1984). They find that the seller's optimal selling mechanism can include
only finite products, despite the socially efficient allocation having a
continuum of products. Like Bergemann, Heumann, and Morris (2022), our
model can also be viewed as an extension of the static screening
problem. Traditionally, screening involves a monopolist offering a menu
of choices to separate consumers based on their private information,
which might seem inconsistent with our informed-seller setup. However,
our analysis focuses on the scenario where the seller cannot use
personalized pricing, making the informed seller's maximization problem
the same as the screening problem when fixing the market segmentation.
This becomes evident in \Cref{sec:discuss-persuasion}, where the
irrelevance result no longer holds and we need to solve a class of
screening probelms with varying segmentations to find the seller's
highest profits.

Bergemann, Brooks, and Morris (2015) analyze a single-product seller and
a continuum of privately informed consumers, assuming the seller can set
different prices for different segments. They characterize the possible
combinations of consumer and seller surplus across all market
segmentations as a triangle. Haghpanah and Siegel (2022) examine when
the surplus triangle of Bergemann, Brooks, and Morris (2015) is
achieveable in the multiproduct scenario. These works differ from ours
in two main aspects. First, they examine all possible segmentations,
while our analysis restricts attention to segmentations that satisfy
seller's incentive-compatibility constraints. Second, they do not
discuss how personalized pricing affects seller profits.

Hidir and Vellodi (2021) assume consumers have private information about
their valuations and can disclose information through cheap talk. They
analyze a horizontally differentiated products setup and characterize
the consumer-optimal market segmentation in equilibrium. The key
tradeoff in equilibirum is that while personalized pricing hurts
consumers by extracting more surplus, it also benefits them by matching
them with more suitable products.

Li and Zhao (2024) analyze an informed seller's joint optimal pricing
and persuasion problem in a credence-good setting, where the consumer
obtains a positive payoff only when consumption exceeds a certain
threshold privately observed by the seller. Li and Zhao (2024) mainly
focus on the linear pricing scenario and characterize the seller's
optimal unit price and disclosure strategy under specific conditions.

\section{Model}\label{sec:model}

There is a seller and a continuum of consumers. The seller (she)
provides \(N\) products, denoted by \(\{ a_1, \dots, a_N \}\). Each
consumer (he) has unit demand, and his action set is
\(A = \{a_0,a_1,...,a_N\}\) where \(a_0\) denotes purchasing nothing.
Our model may also be interpreted such that each \(a_n \ne a_0\)
represents a specific bundle of the seller's products. For example, an
alternative interpretation is that the seller offers only one product,
with \(a_n \in A\) indicating the purchase of \(n\) units of the
product. This interpretation allows for non-unit demand.

The state space is \(Ω \subseteq ℝ^ℓ\) for some positive integer \(ℓ\),
and players' common prior is \(μ_0 \in Δ(Ω)\). The value of \(a \in A\)
to a consumer at state \(ω\) is \(v(ω, a)\), with the value of \(a_0\)
normalized to zero across all states: \(v(ω, a_0) = 0\) for all \(ω\).
At the beginning of the game, the seller posts a price list
\(p: A \to [0,\infty)\) that satisfies \(p(a_0) = 0\) (i.e., the
``price'' for \(a_0\) must be zero). Denote by \(\mathcal{P}\) the set
of price lists.

\paragraph*{Payoffs}\label{payoffs}
\addcontentsline{toc}{paragraph}{Payoffs}

Players are risk-neutral and maximize their expected payoffs. A
consumer's payoff function is \[
u^C (ω, a, p) = v(ω,a) - p(a)
\] for all \(ω \in Ω\), \(a \in A\) and \(p \in \mathcal{P}\). The
seller maximizes her expected profits and her payoff function is \[
u^S(a, p) = p(a) - c(a)
\] where \(c: A \to [0, \infty)\) is the seller's cost function,
satisfying \(c(a_0) = 0\).

\paragraph*{Timing}\label{timing}
\addcontentsline{toc}{paragraph}{Timing}

Timing of the game is as follows: the seller posts a price list
\(p \in \mathcal{P}\); a consumer visits, and the seller privately
observes the state \(ω\), which is drawn from the distribution \(μ_0\);
the seller sends a cheap-talk message\footnote{The message set \(M\) is
  assumed to be sufficiently rich such that \(Δ(A) ∪ Δ(Ω) \subseteq M\).}
\(m \in M\) to the consumer; finally, the consumer chooses \(a \in A\).

The solution concept is perfect Bayesian equilibrium (henceforth,
\emph{equilibrium}). Denote by \(α \in Δ(A)\) a generic consumer mixed
action. We extend the domain of consumer's payoff function \(u^C\) to
include state distributions and action distributions:
\(u^C (μ, α, p) = 𝔼_{μ, α} [u^C (ω, a, p)]\), and analogously for
\(v (ω, a)\) and \(u^S (a, p)\).

\paragraph*{Segmentation}\label{segmentation}
\addcontentsline{toc}{paragraph}{Segmentation}

Through the seller's targeted communication, consumers are divided into
different segments based on their posterior beliefs. For any positive
integer \(K\), call \(\{w_k, μ_k\}_{k=1}^K\) a \emph{(market)
segmentation} if it satisfies \(∑_{k=1}^K w_k μ_k = μ_0\),
\(∑_{k=1}^K w_k = 1\) and \(w_k \ge 0\) for all \(k\). Each segment
\(k \in \{ 1,\dots,K \}\) is characterized by its size \(w_k \in (0,1]\)
and belief \(μ_k\). Denote by \(\mathcal{S}(μ_0)\) the set of all market
segmentations.

Call the pair \((p, \{w_k, μ_k, α_k\}_{k=1}^K)\) an \emph{outcome of the
game}, where \(p\) is the price list and \(α_k \in Δ(A)\) is consumers'
choice at segment \(k\). Our notion of market segmentation is
essentially an interpretation of the distribution of consumer posteriors
in the outcome. This interpretation has been used in the literature of
monopoly information design, such as Bergemann, Brooks, and Morris
(2015), Hidir and Vellodi (2021) and Haghpanah and Siegel (2022). Note
that in these works, consumers are fully informed about their own
valuations, and those within the same segment generally have
heterogeneous valuations. In our model, however, consumers within the
same segment have homogeneous valuations.

\paragraph*{Equilibrium segmentation}\label{equilibrium-segmentation}
\addcontentsline{toc}{paragraph}{Equilibrium segmentation}

A segmentation \(\{w_k, μ_k\}_{k=1}^K\) is an \emph{equilibrium
segmentation under price list} \(p\) if there exist actions
\(\{ α_k \}_{k=1}^K\) such that (i) action \(α_k\) is consumer-optimal
given \(p\) and \(μ_k\) at each segment: \begin{equation}
α_k \in \arg\max_{α \in \Delta (A)} u^C (α, μ_k, p)
  \text{ for all } k \in \{1,...,K \}  \tag{IC-C}
  \label{eq:IC-consumer}
\end{equation} and (ii) the seller's profits from each consumer are
identical, regardless of his segment: \begin{equation}
u^S (α_k,p) = u^S (α_{k'},p) 
  \text{ for all } k, k' \in \{1,...,K\}.  \tag{IC-S}
  \label{eq:IC-seller}
\end{equation} While the two conditions do not directly address the
seller's optimal pricing behavior, they capture the requirements of a
perfect Bayesian equilibrium for a given price list \(p\). Constraints
\eqref{eq:IC-consumer} reflect consumers' incentive-compatibility
constraints at each segment. Constraints \eqref{eq:IC-seller} reflect
the seller's incentive-compatibility constraints. Specifically, since
the seller's payoff is state-independent, cheap-talk communication
requires that she remains indifferent about steering a consumer toward
any action \(α \in \set{α_k}_{k=1}^K\).

\subsection{Seller's highest equilibrium
payoff}\label{sellers-highest-equilibrium-payoff}

Generally, there exist multiple equilibrium segmentations under a fixed
price list. We focus on the scenario where the seller can choose any
equilibrium segmentation following a price list \(p\), and determine the
seller's highest equilibrium payoff. This focus allows us to leverage
the geometric characterization by Lipnowski and Ravid (2020). Below, we
review their main result and introduce some necessary notations.

Denote by \(π (μ ; p)\) the seller's profits without communication under
belief \(\mu\) and price list \(p\); when there exist multiple
consumer-optimal actions, \(π (μ ; p)\) is defined assuming the consumer
chooses a seller-preferred action.\footnote{Let
  \(A^*(μ; p) ≡ \arg \max_{a \in A} u^C (μ, a,p)\) be the set of
  consumer-optimal choices given \(μ\) and \(p\). Then, \(π (μ ; p)\) is
  given by \(\max \{ u^S (a , p) : a \in  A^*(μ ; p)\}\).} For a
real-valued function \(f\) defined on a convex subset of \(ℝ^\ell\),
denote by \(\mathrm{qcav}f\) its \emph{quasiconcave envelope}, the
pointwise lowest quasiconcave function that is everywhere above \(f\).
Lipnowski and Ravid (2020) showed that the seller's highest payoff among
all equilibrium segmentations under \(p\) is
\(\mathrm{qcav}π(μ_0 ; p)\), the quasiconcave envelope of
\(π(\cdot ; p)\) evaluated at the prior \(μ_0\). Consequently, the
seller's highest equilibrium payoff, denoted by \(u^*\), is: \[
u^* = \max_{p \in \mathcal{P}} \mathrm{qcav}π(μ_0 ; p).
\]

However, the above expression requires calculating and comparing the
quasiconcave envelopes for various price lists, and thus is less
practical for computing \(u^*\). Let \[
π^*(μ_0) \equiv \max_{p \in \mathcal{P}} π(μ_0 ; p)
\] be the seller's monopoly profits without communication. Since
\(π^*(μ_0) \ge π(μ_0 ; p)\) for all \(p\), a natural upper bound for
\(u^*\) is \(\mathrm{qcav}π^*(μ_0)\). Note that \(π^* (μ)\) can be
expressed as the upper envelope of \(K+1\) \emph{hyperplanes}:
\begin{equation}
π^* (μ) = \max \{ v(μ, a_k) - c(a_k) : k = 0, 1, ..., K\}.
\label{eq:envelope}
\end{equation} Thus, unlike \(\max_p \mathrm{qcav}π(μ ; p)\), both
\(π^*(μ)\) and its quasiconcave envelope \(\mathrm{qcav}π^*(μ)\) are
straightforward to compute.

\subsection{Personalized pricing}\label{personalized-pricing}

We observe that \(\max_{p \in \mathcal{P}} \mathrm{qcav}π(μ_0; p)\) and
\(\mathrm{qcav}π^*(μ_0)\) represent the seller's highest equilibrium
payoff without and with personalized pricing, respectively. In the setup
described above, personalized pricing is not allowed, and the seller
sets a uniform price list across all segments. If personalized pricing
is allowed, the seller can choose \(K\) price lists
\(\{ p_k \}_{k=1}^K\) alongside a segmentation
\(\{ w_k, μ_k \}_{k=1}^K\), where \(p_k\) is the price list for segment
\(k\). A segmentation \(\{ w_k, \mu_k \}_{k=1}^K\) is said to be an
\emph{equilibrium segmentation under} \(\{ p_k \}_{k=1}^K\) if there
exist actions \(\{ α_k \}_{k=1} ^K\) such that (i) \(α_k\) is
consumer-optimal given \(p_k\) and \(μ_k\) for all
\(k \in \{ 1,...,K \}\) and (ii)
\(u^S (α_k, p_k) = u^S (α_{k'}, p_{k'})\) for all
\(k, k' \in \{ 1,...,K \}\).

Fixing any segmentation \(\{ w_k, μ_k \}_{k=1}^K\), the seller's highest
equilibrium payoff \emph{under personalized pricing} is given by her
\emph{lowest} segment payoff under optimal pricing: \[
\min_{k \in \{ 1,...,K \}} π^* (μ_k).
\] To see this, suppose
\(\min_{k \in \{ 1,...,K \}} π^* (μ_k) = π^* (μ_{\tilde k})\) for some
\(\tilde k\). The seller's incentive-compatibility constraints imply
that her profits cannot exceed \(π^* (μ_{\tilde k})\). Furthermore,
personalized pricing allows the seller to set \(K\) price lists
\(\{ p_k \}_{k=1}^K\) so that her profits in each segment are exactly
\(π^* (μ_{\tilde k})\). Therefore, if personalized pricing is permitted,
the seller will choose a segmentation that maximizes her lowest segment
payoff under optimal pricing. This yields a total payoff of \[
\max_{\{ w_k, μ_k \}_{k=1}^K \in \mathcal{S}(μ_0)} \min_{k \in \{ 1,...,K \}} π^* (μ_k) = \mathrm{qcav}π^*(μ_0),
\] where the quasiconcave envelope is a geometric description of the
seller's highest payoff.\footnote{For details, see Corollary 1 and
  Theorem 2 of Lipnowski and Ravid (2020).}

\subsection{Irrelevance of personalized
pricing}\label{irrelevance-of-personalized-pricing}

Our main result \Cref{prp:main} demonstrates that personalized pricing
is irrelevant to the seller's profits. Specifically, given any
segmentation \(\{ w_i, μ_i\}_{i=1}^K\), we can always find a special
price list \(p^*\) such that \(\{ w_i, μ_i\}_{i=1}^K\) is an equilibrium
segmentation under \(p^*\), and the seller's payoff achives
\(\min_{k \in \{ 1,...,K \}} π^* (μ_k)\), her highest payoff under
personalized pricing.

\begin{prp}

Fixing any segmentation \(\{ w_k, μ_k\}_{k=1}^K \in \mathcal{S}(μ_0)\),
the seller's highest payoff under a uniform price list is given by
\(\min_{k \in \{ 1,...,K \}} π^* (μ_k)\).

\label{prp:main}\end{prp}

\begin{proof}

Let \(\underline{π} ≡ \min_{k \in \{ 1,...,K \}} π^* (μ_k)\). For each
market segment \(k\),~denote by \(φ(k)\)~the index of the product such
that: \[
π^* (μ_k) = v(μ_k, a_{φ(k)}) - c(a_{φ(k)})
\] This induces a map \(φ: \{1,..., K \} \to \{ 1,...,N  \}\). Consider
the price list \(p^*\) defined by: \[
p^* (a_{φ(k)}) = \underline{π} + c(a_{φ(k)}), \text{ for } k = 1,\dots,K
\] and \(p^* (a)\)~is set prohibitively high for
\(a \neq a_{\varphi(k)}\). We verify that \(a_{φ(k)}\) is
consumer-optimal for each \(k\) under \(p^*\). That is,
\(v(μ_k, a_{φ(k)}) - \underline{π} - c(a_{φ(k)}) ≥ v(μ_k, a_{φ(k')}) - \underline{π} - c(a_{φ(k')})\),
which simplifies to: \begin{equation}
v(μ_k, a_{φ(k)}) - c(a_{φ(k)}) ≥ v(μ_k, a_{φ(k')}) -c(a_{φ(k')}). \label{eq:proof-ineq}
\end{equation} Since the left-hand side of inequality
\eqref{eq:proof-ineq} equals \(π^* (μ_k)\), the inequality follows from
equation \eqref{eq:envelope}.

\end{proof}

It immediately follows from \Cref{prp:main} that the upper bound
\(\mathrm{qcav}π^* (μ_0)\) for \(u^*\) is always tight. This result
provides a straightforward method for computing the seller's highest
equilibrium payoff that bypasses consumers' incentive-compatibility
constaints. It can also be equivalently stated as: \[
\max_{p \in \mathcal{P}} \mathrm{qcav}π (μ_0; p) = \mathrm{qcav}\max_{p \in \mathcal{P}} π (μ_0; p) \text{ for all } μ_0 \in Δ(Ω).
\] The economic interpretation is that personalized pricing is
irrelevant to seller's highest equilibirum payoff. We further
demonstrate that there always exists a seller-optimal equilibrium where
each segment has a unique target product.

\begin{crl}

When \(\mathrm{qcav}π^* (μ_0) > π^* (μ_0)\), there exists a price list
\(p^*\), a market segmentation \(\{w_k, μ_k \}_{k=1}^K\) for some
\(K \in [2, N]\), and an into function
\(φ: \{ 1,..., K\} \to \{ 1,...,N \}\) such that for all
\(k \in \{ 1,...,K  \}\):

\begin{enumerate}
\def\labelenumi{\arabic{enumi}.}
\tightlist
\item
  Choosing \(a_{φ(k)}\) is consumer-optimal given \(\mu_k\) and \(p^*\);
\item
  Seller's payoff achieves \(\mathrm{qcav}π^* (μ_0)\) at each segment
  \(k\): \(p(a_{φ(k)} ) - c(a_{φ(k)} ) = \mathrm{qcav}π^* (μ_0)\).
\end{enumerate}

\label{crl:distinct}\end{crl}

\begin{proof}

Our proof of \Cref{prp:main} has established the existence of a
seller-optimal equilibirum outcome where all consumers make
deterministic choices. It suffices to verify that each segment has a
distinct target product in equilibirum. If not, then we can construct a
new outcome through merging: (1) Identify segments where consumers
choose the same product; (2) Merge these segments into a single one, and
consumers in the merged segment choose the same product as before. This
merging process induces a refined seller-optimal equilibrium outcome
where consumers from different segments make different, deterministic
choices.

\end{proof}

\section{Application to credence goods markets}\label{sec:applications}

We apply our characterization of the seller's highest equilibrium payoff
to a credence-good setting, using an \(N\)-by-\(N\) (\(N\) products,
\(N\) states) model that extends Fong, Liu, and Wright (2014).

\subsection{Credence-good setup}\label{credence-good-setup}

There are two players: an expert seller (she) and a client (he). The
client faces a problem but is uncertain about the problem type
\(ω \in  Ω \equiv \{ {ω_1} , \dots ,  {ω_N} \}\). It is common knowledge
that \(ω\) follows the distribution \(\mu_0\) whose support is \(Ω\). If
the problem of type \(ω\) remains unresolved, the client will suffer a
disutility of \(l(ω) \ge 0\). The expert provides \(N\) treatments,
denoted by \(\{a_j\}_{j=1}^N\), which differ in their effectiveness.
Specifically, treatment \(a_j\) can fully resolve problems of type
\(ω_i\) for all \(i \le j\), but cannot resolve problems of type
\(ω_{i}\) where \(i > j\). Treatment \(a_N\) serves as a comprehensive
solution that fully resolves all problems.

\paragraph*{Timing}\label{timing-1}
\addcontentsline{toc}{paragraph}{Timing}

The timing of the game is as follows. First, the expert sets prices for
all treatments. Then, the client visits the expert, who privately
diagnoses the problem type and recommends some treatment \(a_i\).
Finally, the client decides whether to accept or reject the recommended
treatment \(a_i\), where rejection results in no treatment purchase.

\paragraph*{Payoffs}\label{payoffs-1}
\addcontentsline{toc}{paragraph}{Payoffs}

The expert seller's payoff function is \(u^S (a,p) = p(a) - c(a)\) for
\(a \in \{ a_0, ..., a_N\}\), where \(a_0\) represents the client
rejecting the expert's recommendation (i.e., no purchase), \(p(a)\) is
the price for treatment \(a\) with \(p(a_0)=0\) and \(c(a)\) is the cost
of providing treatment \(a\) with \(c(a_0)=0\). To facilitate comparison
with our main model, the client's utility from \(a_0\) is normalized to
zero for all problem types. Thus, the client's payoff function is given
by \(u^C (ω, a,p) = v(ω, a) - p(a)\) where \[
v(ω_i, a_j) = \begin{cases}
l(ω_i) & \text{ if  } j \ge i\\
0    & \text{ otherwise. }
\end{cases}
\] Since the client's payoff from \(a_0\) is normalized to zero,
\(l(\omega)\) can be interpreted as the benefit to the client when the
problem of type \(ω\) is resolved. Assume \(l(ω_i) > l(ω_j)\) and
\(c(a_i) > c(a_j)\) whenever \(i > j\) and \(l(ω_i) - c(a_i) > 0\) for
\(i = 1,2,...,N\).

The solution concept is \emph{perfect Bayesian equilibrium}. We have
assumed that the expert incurs no cost in diagnosing the client's
problem type and that the client is able to observe and verify the
purchased treatment. The latter assumption, which prevents the expert
from charging for one treatment while performing another, is referred to
as the ``verifiability assumption'' in the credence goods literature
(see, for example, Dulleck and Kerschbamer (2006)).

\subsection{Expert's highest equilibrium
payoff}\label{experts-highest-equilibrium-payoff}

The credence-good setup presented here differs from our main model in
two aspects: (a) The expert recommends a specific treatment rather than
communicating through a cheap-talk message; (b) The client can only
accept or reject the recommended treatment, rather than choosing any
\(a \in \{ a_0,...,a_N \}\) as in our main model. Despite these
differences, our characterization of the seller's highest equilibrium
payoff remains applicable.

\begin{prp}

The expert's highest equilibrium payoff is
\(\mathrm{qcav}\pi^*(\mu_0)\).

\label{prp:credence-payoff}\end{prp}

\begin{proof}

We show that \(\mathrm{qcav}\pi^*(\mu_0)\) are both upper and lower
bounds of expert's highest equilibrium payoff. First, expert's highest
equilibrium payoff cannot exceed that of the seller in our main model
with personalized pricing. To see this, observe that any restriction on
the client's choice set following the expert's recommendation can be
replicated in our main model by setting prohibitively high prices for
all treatments except the recommended one. Therefore,
\(\mathrm{qcav}\pi^*(\mu_0)\) is an upper bound of the expert's highest
equilibrium payoff.

Second, consider the seller's problem without personalized pricing in
our main model. \Cref{crl:distinct} establishes the existence of a
seller-optimal equilibirum outcome
\((\{ w_k, \mu_k \}_{k=1}^K, \{ a_{ φ(k) } \}_{k=1}^K, p^*)\) where (i)
consumers at segment \(k\) chooses \(a_{ φ(k)}\) for \(k=1,...,K\) and
(ii) the map \(φ : \{ 1,...,N \} \to \{ 1,...,K \}\) is injective. This
directly translates to an equilibrium outcome in the credence-good setup
where the expert's payoff achieves \(\mathrm{qcav}\pi^*(\mu_0)\): (1)
The expert sets the price list \(p^*\); (2) For all \(k = 1,...,K\), the
client updates his belief to \(μ_k\) after receiving the expert's
treatment recommendation \(a_{ φ(k)}\) and always accepts.

\end{proof}

\subsection{When does communication benefit the
expert?}\label{when-does-communication-benefit-the-expert}

Say that the expert \emph{benefits from communication} when her maximum
equilibirum payoff strictly exceeds her monopoly profits without
communication; that is, \(\mathrm{qcav}\pi^* (μ_0) > \pi^* (μ_0)\). A
natural question is when can communication benefit the expert? We first
look at this question in two special cases.

\begin{eg}

For the binary case, let \(\mu_0= (1 - q_0, q_0)\) where \(q_0\) is the
prior probability weight put on \(ω_2\). When
\(l(ω_1) - c(a_1) = l(ω_2) - c(a_2) > 0\), the expert can extract full
surplus through full revelation and setting \(p(a_1) = l(ω_1)\) and
\(p(a_2) = l(ω_2)\). Thus, she benefits from communication for all
\(q_0 \in (0,1)\). Now assume \(l(ω_1) - c(a_1) < l(ω_2) - c(a_2)\).
\Cref{fig:binary} demonstrates the plots of \(\pi^*(\mu_0)\), depending
on whether there exist prior beliefs such that the client will not
purchase either treatment under optimal pricing with communication, and
of \(\mathrm{qcav}\pi^*(\mu_0)\). While the shape of the expert's profit
function \(\pi^*(\mu_0)\) differs in these two cases, both cases lead to
the same expression of the quasiconcave envelope: \[
\mathrm{qcav}\pi^* (μ_0) = \begin{cases}
l(ω_1) - c(a_1)   & \text{ if } q_0 < \frac {c(a_2) - c(a_1)} {l(ω_2) - l(ω_1)}; \\
l(ω_1) - c(a_2) + (l(ω_2) - l(ω_1)) q_0 & \text{ otherwise}.
\end{cases}
\] Communication benefits the expert if and only if
\(\pi^* (μ_0) < l(ω_1) - c(a_1)\).

\begin{fig}

\includegraphics[width=0.49\linewidth,height=\textheight,keepaspectratio]{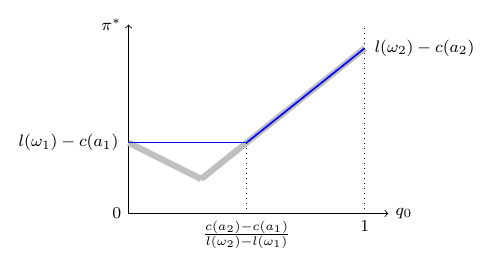}
\includegraphics[width=0.49\linewidth,height=\textheight,keepaspectratio]{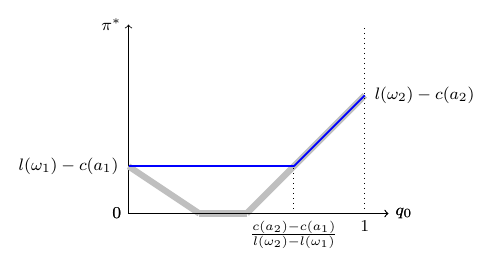}

\caption{Plots of $\pi^* (\mu_0)$ (the blue curve)
  and $\mathrm{qcav}\pi^*(\mu_0)$ (the gray thick curve).
  Left panel: $l(ω_2) \geq \frac{c(a_2) - c(a_1)}{l(ω_1) - c(a_1)} l(ω_1)$. Right panel:
  $l(ω_2) < \frac{c(a_2) - c(a_1)}{l(ω_1) - c(a_1)} l(ω_1)$.}

\label{fig:binary}\end{fig}

\label{eg:binary}\end{eg}

\begin{eg}

For the ternary case, assume
\(l(ω_3) - c(a_3) > l(ω_2) - c(a_2) > l(ω_1) - c(a_1) > 0\).
\Cref{fig:3d} illustrates the surface plots of \(\pi^* (\mu_0)\) and
\(\mathrm{qcav}\pi^*(\mu_0)\) in this case. In the right panel of
\Cref{fig:3d}, the area filled with line patterns indicates the
intersection of the graphs of \(\pi^* (\mu_0)\) and
\(\mathrm{qcav}\pi^*(\mu_0)\). Communication benefits the expert if and
only if \(\pi^* (μ_0) < l(ω_2) - c(a_2)\).

\begin{fig}

\includegraphics[width=0.49\linewidth,height=\textheight,keepaspectratio]{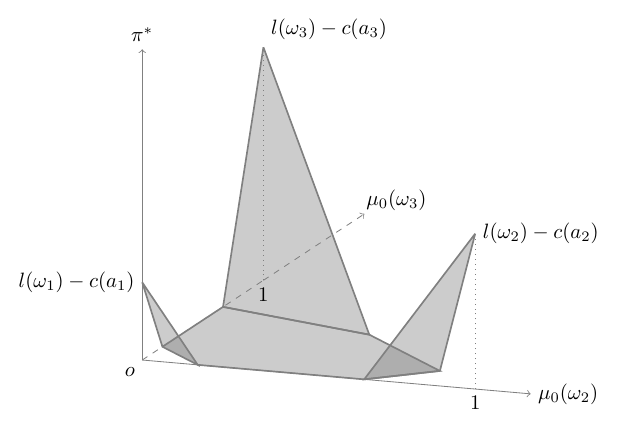}
\includegraphics[width=0.49\linewidth,height=\textheight,keepaspectratio]{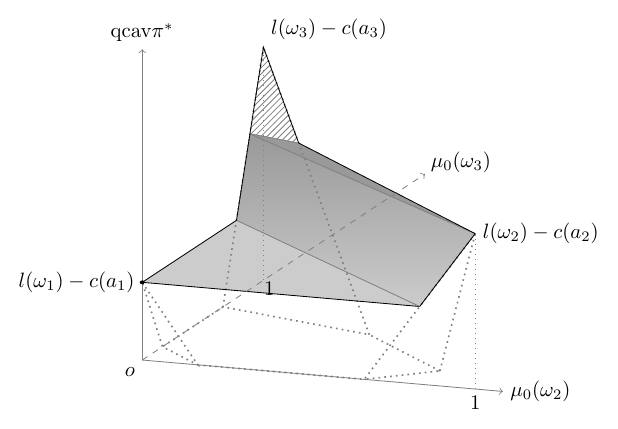}

\caption{Surface plots of $\pi^* (\mu_0)$ (left panel) and $\mathrm{qcav}\,\pi^* (\mu_0)$ (right panel) when $N=3$}

\label{fig:3d}\end{fig}

\label{eg:ternary}\end{eg}

We provide a necessary and sufficient condition for when communication
benefits the expert that generalizes the above examples. Let
\(s_k = l(ω_k) - c(a_k)\) for \(k \in \{ 1,...,N \}\) and denote by
\(s_{(k)}\) the \(k\)-th order element of the set \(\{ s_1,...,s_N \}\)
so that \(s_{(1)} \ge s_{(2)} \ge ... \ge s_{(N)}\).

\begin{prp}

The expert benefits from communication if and only if
\(\pi^* (μ_0) < s_{(2)}\).

\label{prp:when-benefit}\end{prp}

\begin{proof}

See \Cref{app:when}.

\end{proof}

A corollary of \Cref{prp:when-benefit} is that when
\(s_{(1)} = s_{(2)}\), the expert benefits from communication for all
priors \(\mu_0 \in \Delta^o (Ω)\).

\section{Discussion}\label{sec:discuss}

When can an informed seller benefit from personalized pricing? We
provide two examples, demonstrating that personalized pricing can
strictly increase the seller's equilibrium payoff when her
incentive-compatibility constraints are relaxed.

\subsection{When the seller has commitment
power}\label{sec:discuss-persuasion}

The following setup is a specialized case of our main model, except that
the seller can commit to her communication strategy. The seller provides
two products: \(a_1\) and \(a_2\). The state space is
\(Ω = \{ ω_1, ω_2 \}\), and it is common knowledge that
\(\mu_0 (ω_2) = q_0 \in (0, 1)\). The seller's cost function satisfies
\(c(a_1) = 0\) and \(c(a_2) = 1\), and the consumer's value function
\(v(ω, a)\) is defined as in the following table:

\begin{longtable}[]{@{}llll@{}}
\toprule\noalign{}
& \(a_0\) & \(a_1\) & \(a_2\) \\
\midrule\noalign{}
\endhead
\bottomrule\noalign{}
\endlastfoot
\(\omega_1\) & \(0\) & \(1\) & \(1\) \\
\(\omega_2\) & \(0\) & \(2\) & \(4\) \\
\end{longtable}

When personalized pricing is allowed, the seller can extract all
potential social surplus by committing to fully reveal the state,
alongside two price lists \(p_1\) and \(p_2\), for \(ω=ω_1\) and
\(ω=ω_2\) respectively, where \(p_i(a_j) = v(ω_i, a_j)\). One can also
use the concave envelope(Kamenica and Gentzkow (2011)) to describe the
seller's highest profits. For a real-valued function \(f\) defined on a
convex subset of Euclidean space, denote by \(\mathrm{cav}\,f\) its
\emph{concave envelope}, the pointwise lowest concave function that is
everywhere above \(f\). \Cref{fig:persuasion} illustrates
\(\mathrm{cav}\,π^* (μ_0)\), which represents the seller's highest
profits under personalized pricing, and \(π^* (μ_0)\).

\begin{fig}

\pandocbounded{\includegraphics[keepaspectratio]{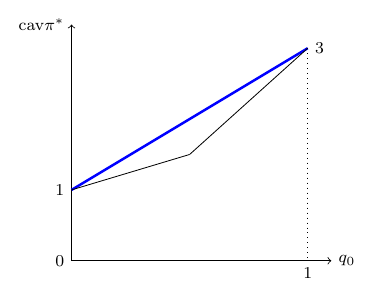}}

\caption{ Example of the seller having commitment power:
  the kinked lines represent $π^* (μ_0)$ and
  the blue line represents $\mathrm{cav}\,π^* (μ_0)$}

\label{fig:persuasion}\end{fig}

\Cref{fig:persuasion} implies that full surplus extraction necessitates
full disclosure from the seller: for any \(q_0 \in (0,1)\), the point
\((q_0, \mathrm{cav}\,\pi^* (μ_0))\) on the blue line can only be
obtained through the convex combination of points \((0, 1)\) and
\((1,3)\). Without personalized pricing, extracting full surplus would
require the seller to fully disclose the state and set a price list
\(p\) satisfying \(p(a_1) = 1\) and \(p(a_2) = 4\). However, this
pricing scheme violates consumers' incentive-compatibility constraints,
as consumers in state \(\omega_2\) would choose \(a_1\) rather than
\(a_2\). Therefore, without personalized pricing, the seller's profits
cannot reach \(\mathrm{cav}\,\pi^* (μ_0)\).

Formally, the seller's highest profits without personalized pricing is
\(\max_{p \in \mathcal{P}} \mathrm{cav}\,π(q_0 ; p)\). We have shown
that
\(\mathrm{cav}\,\max_{p \in \mathcal{P}} π(q_0 ; p) >\max_{p \in \mathcal{P}} \mathrm{cav}\,π(q_0 ; p)\)
for all \(q_0 \in (0,1)\). Computing the seller's highest profits
without personalized pricing is substantially complicated by consumers'
incentive-compatibility constraints. See \Cref{app:commitment} for
detailed derivation.

\subsection{When markets are exogenously
segmented}\label{when-markets-are-exogenously-segmented}

Markets frequently exhibit exogenous segmentation based on observable
consumer characteristics such as age and gender, which are often
correlated with consumers' product valuations. For instance, in
healthcare markets, age significantly influences the probability of
requiring specific treatments. As these exogenous characteristics are
publicly observed by both sellers and consumers, the seller does not
need to maintain indifference across consumers with different
characteristics, potentially making price discrimination profitable for
the seller.

To illustrate this, suppose there are two distinct consumer groups at
the beginning of the game: group 1 with prior belief \(\mu_1\) and size
\(w_1\), and group 2 with prior \(\mu_2\) and size \(w_2\). A real-life
example is the automotive repair market, where drivers of certain
vehicle types may have higher probabilities of requiring costly repairs.
Under personalized pricing, the seller's highest equilibrium payoff is
\(w_1 \mathrm{qcav}\pi^*(\mu_1) + w_2 \mathrm{qcav}\pi^*(\mu_2)\).
However, this payoff may be unattainable under uniform pricing. While
\Cref{prp:main} ensures the existence of price lists \(p_1\) and \(p_2\)
such that \(\mathrm{qcav}\pi^*(\mu_1) = \mathrm{qcav}\pi(\mu_1; p_1)\)
and \(\mathrm{qcav}\pi^*(\mu_2) = \mathrm{qcav}\pi(\mu_2; p_2)\), it
does not guarantee that \(p_1 = p_2\).

\subsection{Concluding Remarks}\label{concluding-remarks}

We conclude by noting that, in reality, another usual source of
personalized pricing arises when consumers possess private information
and when sellers are able to collect data about them to offer
differentiated prices. For instance, recent studies have found that
Didi, a dominant ride-hailing platform in China, conducted a form of
personalized pricing based on the smartphones used by consumers.
Specifically, iPhone users receive fewer subsidies and face higher fares
compared to Android users, with average discounts of RMB 2.07 and RMB
4.12 respectively.\footnote{See Fudan University research reported in
  KrASIA:
  \url{https://kr-asia.com/researchers-took-over-800-trips-using-chinese-ride-hailing-apps-heres-what-they-found}.
  The original report is in Chinese and can be found at
  \url{https://albertlei.github.io/pdf/didi_report_2020.pdf}.}

\section*{References}\label{references}
\addcontentsline{toc}{section}{References}

\phantomsection\label{refs}

\clearpage

\appendix

\section{Appendix}\label{app}

\subsection{\texorpdfstring{Proof of
\Cref{prp:when-benefit}}{Proof of }}\label{app:when}

We want to show that \[
\mathrm{qcav}π^* (μ_0) > π^* (μ_0)
\text{ if and only if }
π^* (μ_0) < s_{(2)},
\] where \(s_{(2)}\) is the second-highest element in the set
\(\{s_1, ..., s_N  \}\). Denote by \(φ(i)\) the index such that \[
s_{(i)} =  l(ω_{φ(i)}) - c(a_{φ(i)}).
\] We first prove the ``only-if'' direction. If \(s_{(2)} = s_{(1)}\),
then \(π^* (μ_0) < s_{(2)} = s_{(1)}\) holds for all
\(μ_0 \in Δ^o (Ω)\). Suppose \(s_{(2)} < s_{(1)}\) and assume by
contradiction that both \(\mathrm{qcav}\,π^* (\mu_0) > π^* (\mu_0)\) and
\(π^*(\mu_0) \ge s_{(2)}\) hold. By \Cref{crl:distinct}, there exists a
price list \(p^*\), a market segmentation \(\{w_k, μ_k \}_{k=1}^K\) and
\(K \ge 2\) distinct products such that the seller's payoff at each
segment achieves
\(\mathrm{qcav}\,π^* (\mu_0) > π^* (\mu_0) \ge s_{(2)}\). This is
impossible because the seller can only achieve a payoff exceeding
\(s_{(2)}\) by selling \(a_{φ(1)}\)---no other product can generate such
a high payoff.

For the ``if'' direction, we first note that the seller can always
achieve the payoff \(s_{(N)}\) by setting the price list \(p\)
satisfying \(p(a_i) = c(a_i) + s_{(N)}\) and fully disclosing the state
to consumers. Thus, we only need to consider cases where
\(π^*(\mu_0) \in [s_{(N)}, s_{(2)})\) and \(s_{(N)} < s_{(2)}\).

Say that \(\{ μ_k \}_{k=1}^{K}\) is a \emph{splitting} of \(μ \in Δ(Ω)\)
if \(μ\) can be written as a convex combination of
\(\{ μ_k \}_{k=1}^{K}\). Our aim is to find a splitting of \(μ_0\) such
that \(π^* (μ_k) > π^* (μ_0)\) for \(k=1,...,K\). If such a splitting
exists, it follows from the Securability Theorem (Theorem 1 of Lipnowski
and Ravid (2020)) that \(\mathrm{qcav}\,π^*(\mu_0) > π^*(\mu_0)\). We
state this observation in the following lemma.

\begin{lmm}

If \(\{ μ_k \}_{k=1}^{K}\) is a splitting of \(μ_0\) and
\(π^* (μ_k) > π^* (μ_0)\) for \(k=1,...,K\), then
\(\mathrm{qcav}\,π^*(\mu_0) > π^*(\mu_0)\).

\label{lmm:split}\end{lmm}

Let \(π_n(μ) \equiv v(μ, a_n) - c(a_n)\) for \(n = 0, 1,...,N\). For any
\(y \in ℝ\), we call \((μ, y) \in ℝ^{N+1}\) a \emph{point}, and refer to
\(y = π_n(μ)\) as the \emph{hyperplane} \(π_n\). The point \((μ, y)\) is
on the hyperplane \(π_n\) if \(y = π_n(μ)\), and two points \((μ, y)\)
and \((μ', y')\) are \emph{never coplanar} if they are not on the same
hyperplane \(π_n\) for all \(n \in \{ 0,1,..., N\}\). Since
\(π^*(μ) = \max \{π_0(μ), π_1(μ), π_2(μ), ..., π_N(μ)\}\), \(π^*\) is
both continuous and convex. Furthermore, the convexity of \(\pi^*\) can
be strengthened in the folllowing sense:

\begin{lmm}

For any two beliefs \(μ_1\) and \(μ_2\), if
\(\pi^* (μ_1) = \pi^* (μ_2) = y\) and \((μ_1, y)\) and \((μ_2, y)\) are
never coplanar, then \(\pi ^* (t μ_1 + (1-t) μ_2) < y\) for all
\(t \in (0,1)\).

\label{lmm:strong-convexity}\end{lmm}

\begin{proof}

Since \(\pi^*\) is convex, \(\pi ^* (t μ_1 + (1-t) μ_2) \le y\) for all
\(t \in (0,1)\). Assume by contradiction that there exists some
\(t' \in (0,1)\) such that \(\pi ^* (t μ_1 + (1-t) μ_2) = y\). Then, it
follows from the convexity of \(\pi^*\) that
\(\pi ^* (t μ_1 + (1-t) μ_2) = y\) for \emph{all} \(t \in (0,1)\). This
implies that \((μ_1, y)\) and \((μ_2, y)\) must both lie on some
hyperplane \(\pi_n\). Contradiction.

\end{proof}

Consider those beliefs \(μ \in Δ(Ω)\) satisfying
\(μ(ω_{φ(N)}) + μ(ω_{φ(1)}) = 1\). That is, \(μ\) is essentially a
two-point distribution whose support is \(\{ω_{φ(1)},  ω_{φ(N)}\}\). Our
analysis of the binary case of the credence-good setup guarantees that
for any \(π^*(μ_0) \in [s_{(N)},  s_{(2)})\), there exists some \(μ_1\)
whose support is \(\{ω_{φ(1)},  ω_{φ(N)}\}\) such that
\(π^*(μ_0) = π^*(μ_1)\). Similarly, there exists another belief \(μ_2\)
whose support is \(\{ω_{φ(2)},  ω_{φ(N)}\}\) such that
\(π^*(μ_0) = π^*(μ_2)\). \Cref{fig:splitting} illustrates \(μ_1\) and
\(μ_2\). Note that by construction, \((μ_1, π^* (μ_1))\) and
\((μ_2, π^* (μ_2))\) are never coplanar.\footnote{That these two points
  are never coplanar depends on the credence-good setup. When only
  states \(ω_{φ(1)}\) and \(ω_{φ(N)}\) are possible, the best-matching
  product for consumer is either \(a_{φ(1)}\) or \(a_{φ(N)}\).
  Therefore, \((μ_1, π^* (μ_1))\) can only be on hyperplane
  \(\pi_{φ(1)}\) or \(\pi_{φ(N)}\), and by construction it is
  \emph{only} on hyperplane \(\pi_{φ(1)}\). Similarly, point
  \((μ_2, π^* (μ_2))\) is only on hyperplane \(\pi_{φ(2)}\).}

\begin{fig}

\includegraphics[width=0.49\linewidth,height=\textheight,keepaspectratio]{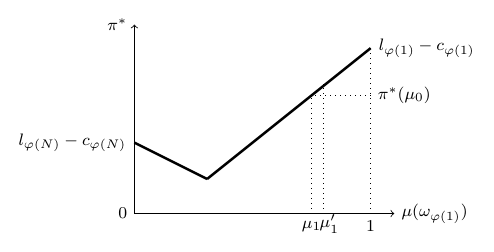}
\includegraphics[width=0.49\linewidth,height=\textheight,keepaspectratio]{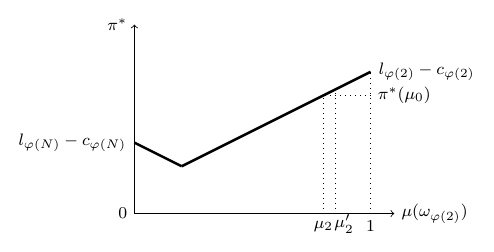}

\caption{Beliefs $μ_1$ and $μ_2$}

\label{fig:splitting}\end{fig}

Since \(μ_0\) is in the interior of \(Δ(Ω)\), there exists another
interior belief \(μ_3\) such that \(\{ μ_1, μ_2, μ_3 \}\) is a splitting
of \(μ_0\). Suppose \(w_1 \mu_1 + w_2 \mu_2 + w_3 \mu_3 = 1\) with
\(w_1 + w_2 + w_3 = 1\) and let
\(\tilde \mu = \frac{w_1}{w_1 + w_2} \mu_1 +  \frac{w_2 }{w_1 + w_2} \mu_2\).
Since \(\pi^*(\mu_1)= \pi^*(\mu_2) = \pi^*(\mu_0)\) and \(\mu_1\) and
\(\mu_2\) are never coplanar, it follows from
\Cref{lmm:strong-convexity} that \(\pi^* (\tilde \mu) < \pi^*(\mu_0)\).
Furthermore, since \(\{\tilde \mu, \mu_3\}\) is a splitting of
\(\mu_0\), the convexity of \(\pi^*(\mu)\) implies
\(\pi^*(\mu_3) > \pi^*(\mu_0)\).

We obtain the desired splitting based on \(\{ μ_1, μ_2, μ_3 \}\) through
perturbation, as illustrated in \Cref{fig:cvx}. Specifically, (1)
\(μ_1'\) is sufficiently close to \(μ_1\) satisfying
\(\pi^*(μ_1') > \pi^*(μ_0)\) and
\(\mathrm{supp}\,μ_1' = \{ ω_{φ(1)},ω_{φ(N)} \}\) (See the left panel of
\Cref{fig:splitting}); (2) \(μ_2'\) is sufficiently close to \(μ_2\)
satisfying \(\pi^*(μ_2') > \pi^*(μ_0)\) and
\(\mathrm{supp}\,μ_2' = \{ ω_{φ(2)},ω_{φ(N)} \}\) (See the right panel
of \Cref{fig:splitting}); (3) Given that \(\{ μ_1, μ_2, μ_3 \}\) is a
splitting of \(\mu_0\), there exists some \(μ_3'\) sufficiently close to
\(μ_3\) such that \(\{μ_1', μ_2',μ_3'\}\) is a splitting of \(\mu_0\).
Since \(π^*\) is continuous and \(\pi^* (μ_3) > \pi^* (μ_0)\), we have
\(\pi^* (μ_3') > \pi^* (μ_0)\) as well.

\begin{fig}

\pandocbounded{\includegraphics[keepaspectratio]{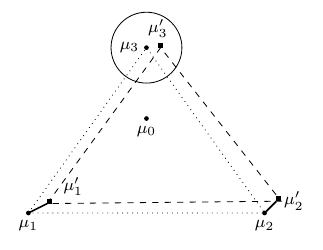}}

\caption{Beliefs $μ_1'$, $μ_2'$ and $μ_3'$ through perturbation}

\label{fig:cvx}\end{fig}

\subsection{Seller's highest payoff with
commitment}\label{app:commitment}

We calculate \(\max_{p \in \mathcal{P}} \mathrm{cav}\,\pi(\mu_0; p)\).
Consumer's belief-based value function is \(v(q, a_1) = q + 1\) and
\(v(q, a_2) = 3q + 1\), where \(q\) is the probability weight put on
product \(a_2\). The seller's highest payoff without communication is \[
\pi^* (q_0) =
  \begin{cases}
    q_0 + 1 & \text{ if } q_0 \leq 1/2;\\
    3q_0 & \text{ if } q_0 > 1/2.
  \end{cases}
\]

\Cref{fig:persuasion} implies that whenever
\(\max_{p \in \mathcal{P}} \mathrm{cav}\,\pi (\mu_0 ; p) > π^*(μ_0)\),
the seller's maximum profits must be achieved via two segments with
posteriors---\(q_1 \in [0, 1/2]\) and \(q_2 \in [1/2, 1]\) satisfying
\(q_1 < q_0 < q_2\)---and some price list \(p\). Once \(q_1\) and
\(q_2\) are fixed, the problem reduces to a standard static screening
problem, and we only need to focus on solutions where both segments are
served by the seller. By standard constrained optimization arguments, we
have \(p(a_1) = q_1 + 1\) and \(p(a_2)\) is determined by the
incentive-compatibity constraints of consumers whose posteriors are
\(q_2\): \[
v(q_2, a_2) - p(a_2) = v(q_2, a_1) - p(a_1) \implies p(a_2) = 2 q_2 + q_1 + 1.
\] Substituting \(p(a_1)\) and \(p(a_2)\) into the seller's profits
yields her reduced payoff:
\(W (q_1, q_2 ; q_0) = \frac{q_2 - q_0 + 2q_0q_2 - q_1 q_2 - q_1^2}{q_2 - q_1}\).
Let \begin{equation}
W^*(q_0) \equiv \max_{q_1 \in [0, 1/2], q_2 \in [1/2, 1]} W (q_1, q_2 ; q_0) \label{eq:reduced}
\end{equation} and we have \[
\max_{p \in \mathcal{P}} \mathrm{cav}\,\pi (\mu;p) = \max \{ W^*(q_0) , π^* (q_0) \}.
\]

Note that \[
\frac{\partial W (q_1, q_2 ; q_0) }{\partial q_2} = \frac{(q_0 - q_1) (1 - 2 q_1)}{(q_2 - q_1)^2} > 0.
\] Thus, \(q_2^* = 1\) in the solution to the problem
\eqref{eq:reduced}. Substitute \(q_2^* = 1\) into \(W (q_1, q_2 ; q_0)\)
and the first-order condition with regard to \(q_1\) yields: \[
q_1^* =
  \begin{cases}
    1 - \sqrt{1 - q_0} & \text{ if } q_0 \leq 3/4 \\
    1/2 & \text{ if } q_0 > 3/4.
  \end{cases}
\] Substitute \(q_2^*\) and \(q_1^*\) into \(W (q_1, q_2 ; q_0)\) yields
\(W^*(q_0) = 3 - 2 \sqrt{1 - q_0}\) when \(q_0 \leq 3/4\) and otherwise
\(W^*(q_0) = 2 q_0 + 1/2\). Therefore, the seller's maximum payoff
without personalized pricing is \[
\max \{ W^* (q_0), \pi^* (q_0)\} =
  \begin{cases}
    3 - 2  \sqrt{1 - q_0}  & \text{ if } q_0 \leq 5/9; \\
    3q_0 & \text{ if } q_0 > 5/9 .
  \end{cases}
\]

\end{document}